\documentclass[journal]{IEEEtran}
\IEEEoverridecommandlockouts
\usepackage{cite}
\usepackage{amsmath,amssymb,amsfonts}
\usepackage{algorithmic}
\usepackage{graphicx}
\usepackage{textcomp}
\usepackage{xcolor}
\usepackage{comment}
\usepackage[caption=false]{subfig}
\usepackage{url}

\usepackage{soul}
\usepackage[english]{babel}
\usepackage{rotating}
\usepackage{array}
\usepackage[switch]{lineno}
\usepackage{soul}

\usepackage{array}
\newcolumntype{L}[1]{>{\raggedright\let\newline\\\arraybackslash\hspace{0pt}}m{#1}}
\newcolumntype{C}[1]{>{\centering\let\newline\\\arraybackslash\hspace{0pt}}m{#1}}
\newcolumntype{R}[1]{>{\raggedleft\let\newline\\\arraybackslash\hspace{0pt}}m{#1}}

\def\BibTeX{{\rm B\kern-.05em{\sc i\kern-.025em b}\kern-.08em
    T\kern-.1667em\lower.7ex\hbox{E}\kern-.125emX}}

\title{The Ethics of AI in Games}

\author{
    \IEEEauthorblockN{David Melhart, Julian Togelius, Benedikte Mikkelsen, Christoffer Holmgård, Georgios N. Yannakakis}\\
    \IEEEauthorblockA{\emph{modl.ai}\\
    Copenhagen, Denmark \\
    david@modl.ai, julian@modl.ai, benedikte@modl.ai, christoffer@modl.ai, georgios@modl.ai }
}

\begin{document}

\maketitle

\begin{abstract}
Video games are one of the richest and most popular forms of human-computer interaction and, hence, their role is critical for our understanding of human behaviour and affect at a large scale. 
As artificial intelligence (AI) tools are gradually adopted by the game industry a series of ethical concerns arise. Such concerns, however, have so far not been extensively discussed in a video game context. Motivated by the lack of a comprehensive review on the ethics of AI as applied to games, we survey the current state of the art in this area and discuss ethical considerations of these systems from the holistic perspective of the \emph{affective loop}. Through the components of this loop, we study the ethical challenges that AI faces in video game development. \emph{Elicitation} highlights the ethical boundaries of artificially induced emotions; \emph{sensing} showcases the trade-off between privacy and safe gaming spaces; and \emph{detection}, as utilised during in-game \emph{adaptation}, poses challenges to transparency and ownership. This paper calls for an open dialogue and action for the games of today and the virtual spaces of the future. By setting an appropriate framework we aim to protect users and to guide developers towards safer and better experiences for their customers.
\end{abstract}


\section{Introduction}

Video games are key to our understanding of human behaviour due to their vast popularity, the multi-modal ways players can interact with them, and the various ways games can express emotion and adapt to a player’s style. Even though values such as transparency, trustworthiness and responsibility are core aspects of ethical systems in other domains, video games present unique challenges in terms of ethics. 
Dark patterns in game design \cite{zagal2013dark}, predatory monetisation strategies \cite{king2019unfair}, and the black-box nature of games hinder transparency \cite{mikkelsen2017ethical} and raise several ethical concerns. These issues are far-reaching from game design and development \cite{cook2017ethical,seif2020data} to societal impact and research ethics \cite{cook2021social}. 

\begin{figure}[!tb]
    \centering
    \includegraphics[width=1\linewidth]{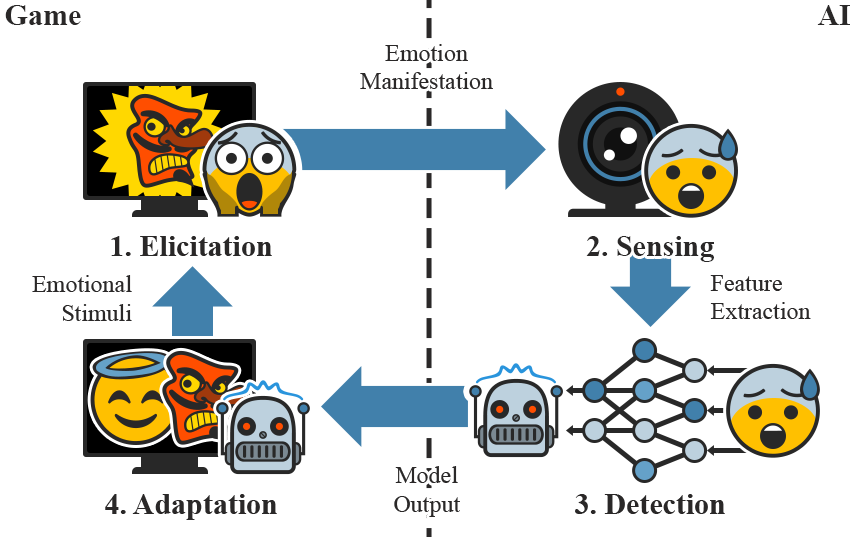}
    \caption{The \emph{Affective Game Loop} \cite{yannakakis2014emotion}. The loop relies on the game's parameter space to \emph{elicit} an emotional response. This response is \emph{sensed} by an AI model that \emph{detects} change(s) in the player's emotional state. The output of the affect model can be used to \emph{adapt} the game content and generate a new set of stimuli for the player.}
    \label{fig:affective_loop}
\end{figure}

\begin{figure}[!tb]
    \centering
    \includegraphics[width=1.0\linewidth]{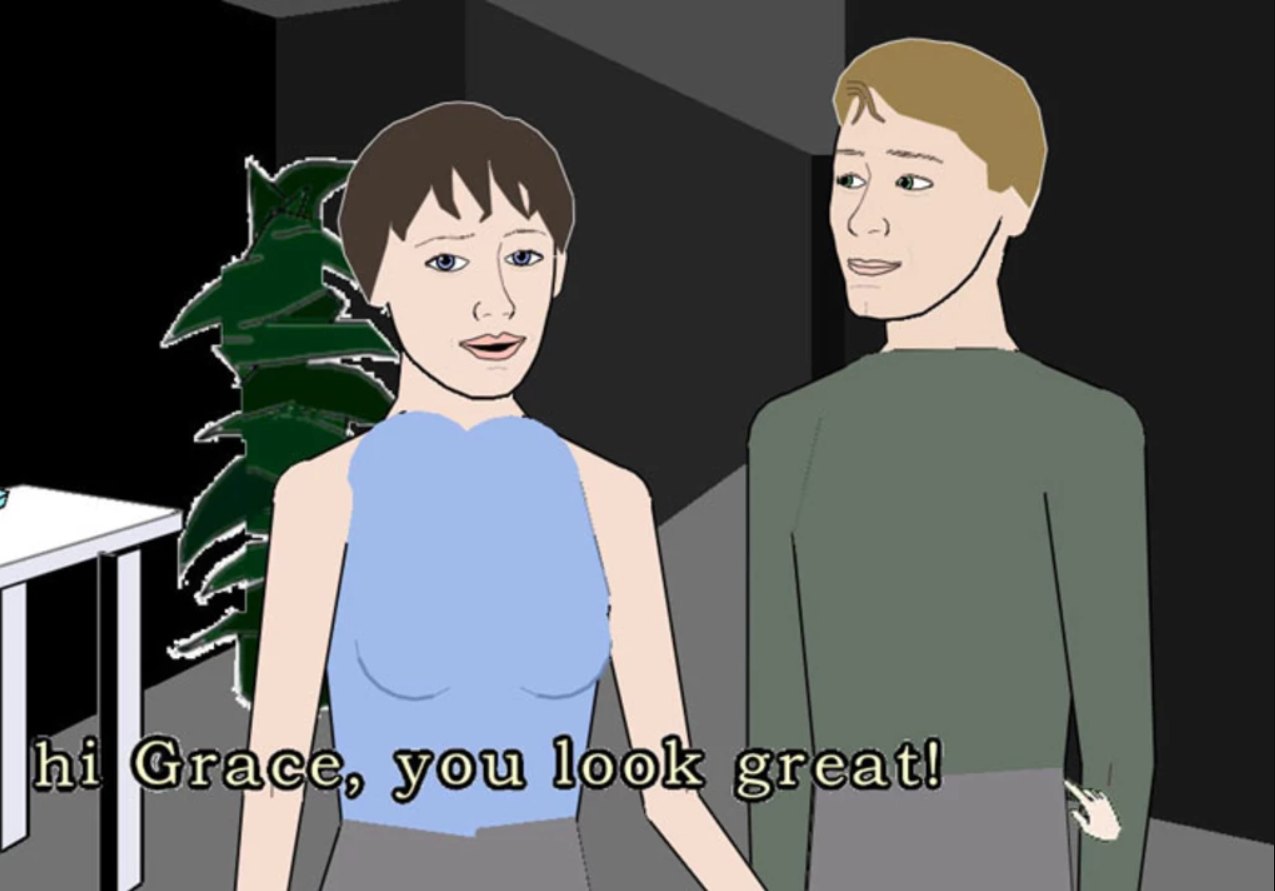}
    \caption{In \emph{Façade} (Procedural Arts, 2005), the player can interact with the game's agents through free-form text. The underlying AI responds to the player input based on its semantic and emotional content.}
    \label{fig:facade}
\end{figure}

\begin{figure}[!tb]
    \centering
    \includegraphics[width=1.0\linewidth]{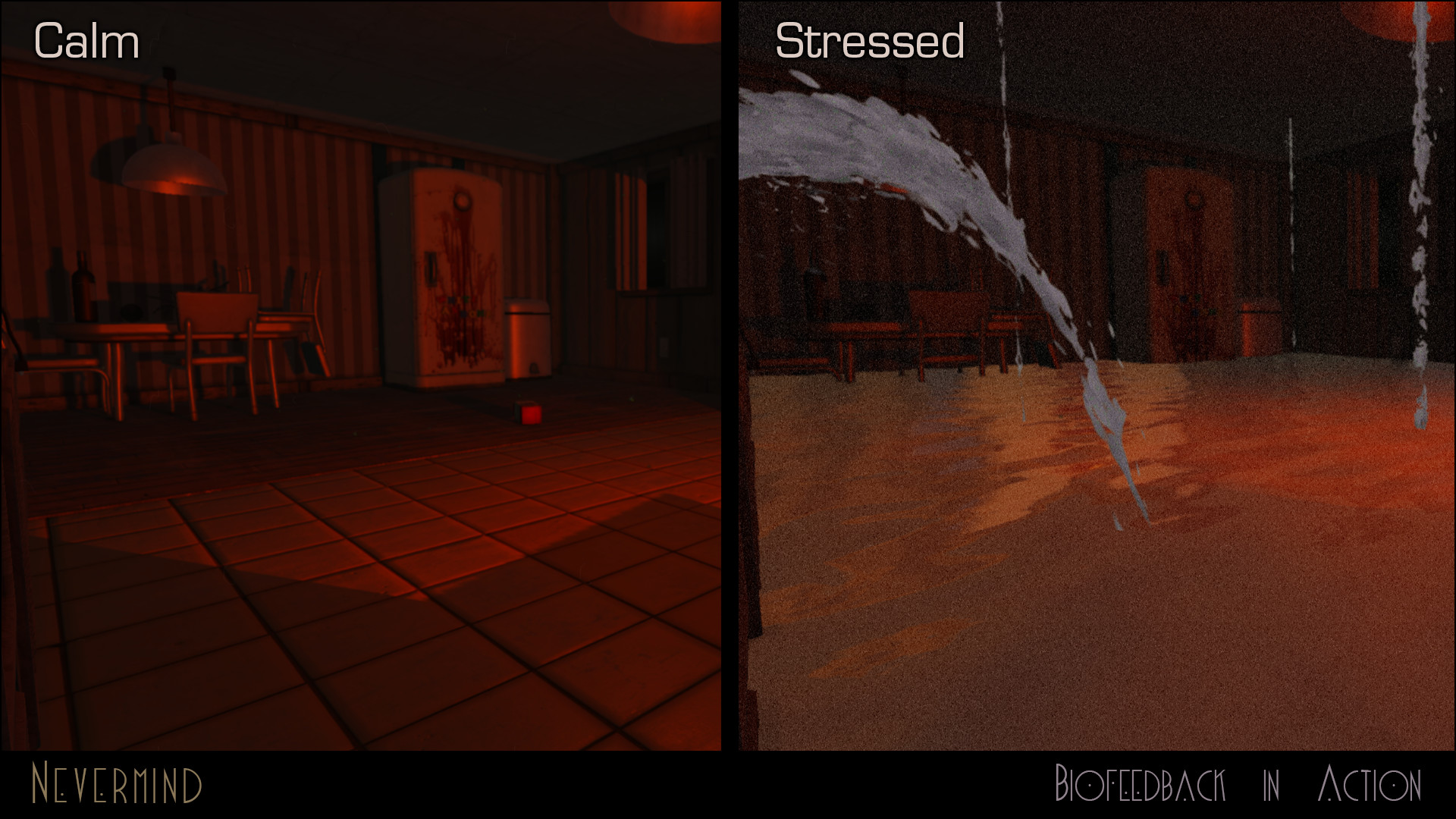}
    \caption{In \emph{Nevermind} (Flying Mollusk, 2016), the player explores dream-like horror environments. The game content is adjusted based on the player's emotional state by introducing more dangers as the player's stress level increases.}
    \label{fig:nevermind}
\end{figure}

In this survey paper, we aim to address the ethical considerations of game AI tools and methods through the lens of \emph{affective computing}. In particular, we focus primarily on \emph{player modelling} \cite{yannakakis2013player} as a field of game research that considers the aggregation, simulation \cite{holmgaard2014evolving}, and understanding of gameplay and user experience in games. We, thus, structure the discussion of AI ethics in games around the \emph{affective game loop} \cite{yannakakis2014emotion} (see Fig.~\ref{fig:affective_loop}). The affective game loop describes the relationships between emotion expression, elicitation, detection, prediction, and subsequent reaction. It presents a complex game system which facilitates these processes and adapts to the user's emotional response. This loop can assist AI systems to generate personalised aspects of games such as agent behaviour, levels and images \cite{yannakakis2011experience} or guide an orchestration process \cite{liapis2018orchestrating,togelius2016general} across creative facets such as text, levels and visuals. 
The concept of the affective game loop has been explored thoroughly in academia
\cite{yannakakis2009real,shaker2010towards,van2016deep,villani2018videogames,melhart2021towards}. Meanwhile, the adoption of affect-driven adaptation systems in games has been gradual over the last twenty years; indicative yet representative examples include \emph{Façade} (Procedural Arts, 2005)---see Fig.~\ref{fig:facade}---and \emph{Nevermind} (Flying Mollusk, 2016)---see Fig.~\ref{fig:nevermind}. 

\begin{table*}[!tb]
    \centering
    \caption{Aspects of the Affective Game Loop with their associated AI Virtues (introduced by Bostrom and Yudkowsky \cite{bostrom2014ethics}); major pitfalls; and positive initiatives upholding AI Virtues and benefiting end-users.}
    \begin{tabular}{|l||c|L{2cm}|L{6cm}|L{5cm}|}
        \hline
        \textbf{Affective Loop} & \textbf{Section} & \textbf{AI Virtues} \cite{bostrom2014ethics} & \textbf{Major Pitfalls} & \textbf{Positive Initiatives} \\
        \hline
        \hline
        \textbf{Elicitation} & \ref{sec:elicitation} & Responsibility, Auditability & Predatory monetisation, dark design patterns \cite{zagal2013dark}, and generating harmful content & Governments tightening regulation around predatory monetisation practices \cite{subhan_2022}; Prolific designers taking a critical stance against dark patterns (e.g. Six to Start CEO Adrian Hon \cite{hon2022you}) \\
        \hline
        \textbf{Sensing} & \ref{sec:sensing} & Transparency, Incorruptibility & Lack of transparency in data inferred by AI systems & Ubisoft's ML models sensing for toxic behaviour in \emph{For Honor} (2017) \cite{canossa2021honor} \\
        \hline
        \textbf{Detection} & \ref{sec:detection} & Transparency, Auditability, Predictability & AI systems overfitting to skewed populations and perpetuating harmful historical biases & Game studios sharing datasets with the academic community (e.g. EA and Nintendo \cite{johannes2021video}, Ubisoft \cite{melhart2019your} and Riot Games \cite{monge2022effects}).\\
        \hline
        \textbf{Adaptation} & \ref{sec:adaptation} & Responsibility, Transparency, Incorruptibility & Unclear chain of responisibiliy and ownership of data and output of human-AI co-creation & Microsoft's \emph{Xbox Transparency Reports} showing actions taken in content moderation\footnote{\url{https://www.xbox.com/en-GB/legal/xbox-transparency-report}}.  \\
        \hline
    \end{tabular}
    \label{tab:stucture}
\end{table*}

The paper is structured as follows. After an overview of related literature (Section \ref{sec:related_work}), we discuss the ethical dimensions of game AI through the phases of the \emph{affective game loop}---for a detailed structure of our survey see Table~\ref{tab:stucture}. In particular, Section~\ref{sec:elicitation} covers aspects of \emph{elicitation} and how dark patterns are used to manipulate and reduce the players' emotional agency in harmful or exploitative ways; Section~\ref{sec:sensing} takes a thorough look at \emph{sensing} and issues related to the tradeoff between privacy and control, and malicious action in games; Section~\ref{sec:detection} discusses affect \emph{detection} and the complexities of transparency in limited information systems such as games; and finally Section~\ref{sec:adaptation} reflects on questions of data and model ownership during the affect-driven \emph{adaptation} phase. The paper ends with a discussion on several other issues related to game AI ethics including AI algorithmic biases, compute fairness, and in-game toxicity and violence.

\section{Related Work}\label{sec:related_work}

This section reviews the literature on ethics research and ethical frameworks in AI and games research.

\subsection{Ethics in Artificial Intelligence}

Ethics has been a constant challenge in the field of AI fuelled by academic and practical interest into the governance of autonomous systems and public anxiety towards data-driven black-box infrastructures \cite{bryson2011just,yu2018building}. Ethical frameworks have been developed to address these anxieties, generally aiming to provide guidelines for creating beneficial, transparent, and trustworthy applications.

One of the most popular ethics frameworks applied in AI consists of the virtues of \emph{Responsibility}, \emph{Transparency}, \emph{Auditability}, \emph{Incorruptibility}, and \emph{Predictability} introduced by Bostrom and Yudkowsky \cite{bostrom2014ethics}. The \emph{responsibility} of AI algorithms refers to their clear oversight on the chain of responsibility as the output of the algorithm can be attributed to either individuals or organisations \cite{crowley2019toward}. As we will discuss in Section~\ref{sec:elicitation} and~\ref{sec:adaptation}, a clear chain of responsibility is often lost between the original data, the inferred models, and large-scale ensemble architectures using third-party AI. 

\emph{Transparency} is one of the more complex ethical virtues and cornerstones of AI Trustworthiness \cite{larsson2020transparency,thiebes2021trustworthy}. On the one hand, it can refer to a kind of \emph{algorithmic transparency} that promotes
AI decision-making processes that are explainable \cite{das2020opportunities} and clearly understood by their users \cite{crowley2019toward}. 
On the other hand, it can refer to a \emph{systemic transparency and openness} of AI-powered applications; that is legal access to AI infrastructures themselves \cite{larsson2020transparency}.
As we will show in Section~\ref{sec:sensing}, Section~\ref{sec:detection}, and Section~\ref{sec:adaptation}, the industry has a troubled relationship with transparency, with many companies not disclosing their use of player models to gain further insights from user data. While transparency in relation to direct data-collection is clear, ``inferred'' information such as computational models and their output is much less protected by legal frameworks \cite{wachter2019right}. 

\emph{Auditability} implies that the correctness of the output of AI systems should be verifiable by a third party. As we discuss in Section~\ref{sec:elicitation} and Section~\ref{sec:detection}, auditability and general transparency is a serious blind spot of the video games industry. 
Although some of this blind-spot can be attributed to the inherent opacity of AI systems \cite{beckman2022artificial}, there is a definite limitation raised by legal opacity restricting access to AI architecture and training data by external auditors \cite{liu2021trustworthy}.

\emph{Incorruptibility} means that the system is robust against manipulation. Even though the obfuscation of datasets, algorithms, and their output definitely provides some level of protection, obfuscation is fundamentally clashing with the principle of transparency. Due to their interactivity games are under constant siege by malicious users, however, their corruption is not necessarily an outside force. As Gebru points out, AI bias tends to exacerbate the sociopolitical and socioeconomic disparities in our society as they perpetuate inherent biases of the creators of AI models and our social reality \cite{gebru2020race}. As we discuss in Section~\ref{sec:sensing}, while one of the primary goals of applied AI in the game industry is to increase the robustness of systems against external attacks, there is much less discussion and transparency about the inherent bias in the employed AI systems.

Finally, \emph{predictability} refers to self-consistent AI outputs and algorithmic behaviour. Predictability is a less prominent yet important aspect of AI ethics, which aims to push applications towards a more reliable and fair implementation \cite{crowley2019toward}. Predictability goes a long way towards eliminating AI bias, which we detail in Section~\ref{sec:detection}.
The aforementioned virtues are being understood as the cornerstones of AI ethics and solidified \cite{mikkelsen2017ethical,vakkuri2019ethically}---in some shape or form---in the \emph{IEEE Ethically Aligned Design Guidelines} \cite{ieee2019ethics}, the \emph{Humane AI Ethical Framework} \cite{crowley2019toward} and the newly emerging concept of \emph{Trustworthy AI} \cite{floridi2019establishing,thiebes2021trustworthy}, which also plays a fundamental role in the new \emph{Ethical Guidelines for Artificial Intelligence of the European Union} \cite{smuha2019eu}. 

A recent meta-review by Yu et al. \cite{yu2018building} of the AAAI, AAMAS, ECAI and IJCAI conferences mapped out the field of Ethics in AI (EAI) and identified four major areas under this domain. The first category is research focusing on leveraging AI techniques to \emph{explore questions of ethics} faced by humans. The second and third categories focus on internal \emph{decision-making frameworks} for AI agents acting either as individual units or collectives. Finally, the last category focuses on \emph{ethics in human-computer interactions}. For a complete review of all these avenues of research we refer to Yu et al. \cite{yu2018building}; here we focus only on the latter category as it is the most relevant to the domain and purposes investigated in this paper. As positioned by Yu et al. \cite{yu2017towards,yu2018building} and echoed by the larger research \cite{dignum2018ethics,vakkuri2019ethically,mikkelsen2017ethical} and policy making \cite{ieee2019ethics,floridi2019establishing} communities, ethical HCI systems should conserve the autonomy of humans, be beneficial to the user, and minimise underlying risks. Summarising this sentiment in relation to affective computing, the \emph{IEEE Ethically Aligned Design Guidelines} explicitly state:

\begin{quote}
    ``To ensure that intelligent technical systems will be used to help humanity to the greatest extent possible in all contexts, autonomous and intelligent systems that participate in or facilitate human society should not cause harm by either amplifying or dampening human emotional experience'' \cite[page~6]{ieee2019ethics}. 
\end{quote}

Beyond the scope of emotional autonomy, however, there is also the question of transparency and autonomy in human-AI interaction in general. Rovatsos raises the issue of the general distrust towards machines and whether it is ethical for an AI system to conceal itself \cite{rovatsos2019we}. Although it can be easy to consider total transparency as the most ethical, the issue is more complex. A new ethical conundrum emerges when we consider that in some human-computer interactions, a lack of transparency can improve the efficacy of the system \cite{ishowo2019behavioural}. If this is true, wouldn't the performance drop---that was induced by increased transparency---hurt the user in the long run? Would in this situation total transparency take away from the user's autonomy? On the other hand, could an opaque system even present fair choices to the user? These questions---raised by Rovatsos \cite{rovatsos2019we}---presuppose a benevolent system. However, AI is not always designed to be benevolent. Perhaps the most striking example of this is lethal autonomous weapon systems, which are designed to kill humans without considerable oversight \cite{russell2015ethics}. Even though real-life killing robots might seem to be removed from the domain of games, pushing a military agenda and aiding both recruitment and research has never been far from video games \cite{cook2021social}. And as we discuss at many points in this paper neither is emotional exploitation nor psychological manipulation.

AI researchers from the fields of computer science, engineering, robotics, medicine, games, and more are calling for stronger regulations on exploitative and harmful AI and a push for benevolent AI applications \cite{russell2015ethics,russell2019human,cook2021social}. Their fears are not unfounded as current research and industrial application of AI are more than capable of exploiting and harming humans en masse, from social engineering \cite{hagendorff2020ethics}, through psychological manipulation \cite{king2019unfair} and exacerbating existing socioeconomic disparities to physical harm \cite{russell2015ethics,cook2021social}.

\subsection{AI Ethics in Game Research}

AI ethics in game research is a fairly under-researched area. The handful of papers existent in the literature focus mainly on player modelling \cite{mikkelsen2017ethical,zhu2021open}, ethical development practices \cite{seif2020data,cook2017ethical} and ethical practices in research \cite{cook2021social}. In contrast, more work has been carried out on games as ethical systems \cite{sicart2011ethics} and the outcome of responsibility of game design \cite{zagal2013dark}. 

In a review of the field of player modelling, Mikkelsen et al. provided an overview of emerging ethical issues \cite{mikkelsen2017ethical}. In their analysis relying on the framework laid out by \cite{bostrom2014ethics}, Mikkelsen et al. identified a number of areas of concern from monetisation through content management to dynamic adaptation and privacy. Most issues emerging in these areas are connected to the lack of \emph{transparency} and \emph{auditability} of computer models, especially in industrial settings. One solution to the lack of transparency and interpretability is offered by the field of \emph{explainable AI}  \cite{zhu2018explainable,arrieta2020explainable}. A possible avenue for adapting explainable AI is through \emph{open player models} \cite{zhu2021open}, which are based on Open Learner Models applied to games \cite{hooshyar2020transparent}. Open player models incorporate an explanatory module into a given AI application which gives clear feedback to the user on the behaviour and predictions of the model. As the module providing transparency is removed from the main pipeline of the algorithm, in principle open player models can be cost-effective to implement in existing systems as well. Nevertheless, although explainable AI principles can help build more transparent systems in theory, the practical application of such frameworks appears to be challenging. Black-box algorithms such as deep learning neural networks are very popular in data science due to their performance, and they are notoriously hard to explain and interpret---despite advances \cite{arrieta2020explainable}. On the other hand, effective white-box systems are still an open challenge to the field \cite{zhu2018explainable} and might not even be sustainable or desirable from a business perspective as the games industry is known to treat datasets, data-processing pipelines, and AI models as strictly-kept trade secrets.

Beyond the concerns of transparency there is an alarming issue of intentionally harmful usage of AI models that exploit addiction and irresponsible spending habits \cite{mikkelsen2017ethical,seif2020data}. Despite a growing concern against aggressive and deceptive monetisation techniques---often aimed at children---there is still a lack of legal and practical frameworks that are capable of addressing such issues \cite{king2019unfair}. King et al. \cite{king2019unfair}, for instance, examined 13 different patents connected to video game monetisation and found that almost all of them relied on the exploitation of the players' data to optimise the delivery and timing of ads and purchase offers. They note that with the expansion of AI methods it is expected that such systems will become more sophisticated and ubiquitous in the future, making the issue of ethics in player modelling more pressing than ever.

\subsection{State of AI Ethics in Practice}

Although ethical frameworks have been developed to provide guidance, the lack of specificity often leads to a small scale of adoption. If we look at the core issue of \emph{transparency}, which is also often required for the assessment of other components of AI trustworthiness, we find that both affective computing and games applications are lagging behind \cite{kroger2023surveilling}. This is true despite the issue of transparency being propped up by more robust legal frameworks than many other components of ethical AI. 
In the European Union, the General Data Protection Regulation (GDPR) \cite{voigt2017eu} is meant to give a legal framework and transparency to data handling (also in an AI context). However, a review of \emph{serious games}---games developed for healthcare, educational, hiring, or other non-entertainment purposes---found that two years after the adoption of GDPR, it has had little to no effect on the research community \cite{jost2020two}. 
Similarly, in a recent exploration of affective computing through the lens of GDPR laws Hauselmann found that the field faces serious issues in terms of \emph{transparency}, \emph{responsibility}, and \emph{predictability} \cite{hauselmann2021fit}. Hauselmann highlights the delicate nature of emotional data as something that is not necessarily protected under current legal frameworks but extremely personal to the users. However, the question of emotional data is further complicated by the fact that while user behaviour is relatively easy to observe and record, emotional data is often extracted through means of peripheral signals and machine learning. In this sense affective data is inferred and not observed \cite{wachter2019right}; as a result, the majority of affective computing applications appear to be inherently opaque.
As there should be a right to an accurate portrayal of personal data, inaccurate predictors might infringe on the personal rights of users. This is hard to prove, however, as these models are often difficult to audit. This phenomenon is amplified because there are fewer practical concerns for inaccurate models up to a certain degree. Often even if a user is profiled inaccurately, an imperfect prediction can still be used to great effect in an adaptive system \cite{mccrea2017absolute}. Moreover, commercial applications often safeguard their models as trade secrets or cannot handle the constraints and overhead of implementing ethical safeguards on a fundamental level. 

The above examples focus predominantly on the research community; in the games industry, the problem of transparency can be even more prevalent. More often than not, users are unaware of the data collected and inferred by algorithms. As Kr{\"o}ger points out, data collection in games is generally made invisible to the players as it is ``woven into a game’s environment'' \cite{kroger2023surveilling}. Given this opaqueness and a blas{\'e} attitude of users towards---what they perceive as---anonymous play, it is questionable to which extent regulations such as the aforementioned GDPR could reasonably be upheld.
A thorough review of five companies by Vakkuri et al. \cite{vakkuri2019ethically} revealed that even though developers might consider ethics as an important question, they have little to no tools to address it in a systematic manner. Mitigation of ethical risks in AI systems thus becomes low-priority and generally addressed in a post- and ad-hoc manner, if at all. Reviews of game industry applications that span from player modelling \cite{mikkelsen2017ethical,zhu2021open}, to data-driven game development \cite{seif2020data}, and to procedural content generation \cite{cook2017ethical} reveal a similar pattern. Because the games industry is a fast-moving field with growing pressure on producing more and more content with the advent of live-service games, the application of ethical frameworks to AI in games---including but not limited to player modelling---remains an after-thought without clear ways to integrate the mitigation of ethical problems into existing industry pipelines. 

\section{Elicitation -- Boundaries of Artificially Induced Emotions}\label{sec:elicitation}

We start examining the affective game loop from the \emph{Elicitation} phase. Doing so we are faced with the ethical boundaries of artificially induced emotions. Although inherently personal and subjective, emotions do not enjoy legal protection to the same extent as other personal data \cite{hauselmann2021fit}. The core issues we encounter in this area are ownership and autonomy over one's own emotions. The so-called \emph{dark design patterns} \cite{zagal2013dark} have been used in games to compel players' behaviour through affective manipulation and with the advent of big data analysis and machine learning, there is a potential for a new wave of dark design patterns \cite{king2019unfair}. As games are often marketed towards children, the ethical side of the emotion elicitation in games, their use and their goal have to be considered. Importantly, the challenge of dark design patterns is core to game design principles but not necessarily to the AI algorithm associated with a game. One should thus take a dive into the problematic ethical aspects of the game design prior to examining the role of AI within a particular game.

While a few years ago \emph{loot boxes}---i.e. virtual items that can be redeemed for other random items that provide some value to the players \cite{zendle2018video}---made waves \cite{drummond2020loot,gibson2022relationship,hodge2022box}, the new monetisation technique sweeping across the industry is the \emph{battle pass} or \emph{season pass} system. Unlike previous iterations of \emph{premium subscriptions}, \emph{in-game currencies}, \emph{downloadable content} packs, \emph{loot boxes}, and \emph{gated progression}, the \emph{battle pass} system does not promise any immediate tangible reward to players. Instead, players buy into access to time-limited content updates, which they still have to unlock in-game within a given time frame \cite{petrovskaya2020battle}. This type of monetisation reformulates the value proposition of online games and shifts the focus from commodities to services \cite{joseph2021battle}. While the loot boxes of yesteryear were designed to operate on the same psychological buttons as gambling \cite{drummond2020loot,gibson2022relationship}, emerging battle pass systems build more on a \emph{feeling of missing out} \cite{petrovskaya2020battle} and societal pressure \cite{king2020fortnite}. In many modern online games---such as \emph{Fortnite} (Epic Games, 2017), \emph{Apex Legends} (Respawn Entertainment, 2019), \emph{Fall Guys} (Mediatonic, 2020) and \emph{Overwatch 2} (Blizzard Entertainment, 2022)---these monetisation practices often coalesce into a virtual storefront, where in-game currencies can be bought for real money, then spent on single purchase upgrades and battle passes alike.

Where player modelling techniques can make loot boxes, battle passes, and other similar monetisation techniques more concerning is the ability to target people more prone to spending. Predictive models have already been in place for years in the industry for the estimation of churn \cite{perianez2016churn} to keep track of players lost and their velocity through a game. Similar models, however, can also be used to find and target potential excessive spenders---often called ``whales'' in the industry\cite{chen2018customer}. Affective computing models estimating the users' emotional state can be used for the targeted and timed delivery of ads and promotional offers to maximise user spending. Affective modelling methods, however, can also help build more responsible systems that detect a risk of a problematic behaviour. Out of the psychosocial aspects of addiction (salience, mood modification, tolerance, withdrawal symptoms, relapse, conflict) \cite{griffiths2005components}, affective computing methods could be especially useful to pick up on mood modification and tolerance (as a diminishing affective response) at the very least and flag users as at-risk consumers. In the same way, machine learning models based on affective and behavioural feedback \cite{xi2019latent} are deployed to target monetisation and retain consumers, they can also be used to deploy ``precision psychiatry'' \cite{mak2019applications}. Most notably, EA was accused to leverage their patented dynamic difficulty adjustment system to push players to spend more money on loot boxes \cite{valentine2020ealawsuit}. Even though the case was dismissed and EA swore to uphold ``fair play'' in their online games \cite{batchelor2021ealawsuit}, this indicative example goes to show how algorithms can be used to elicit emotions that influence players to act against their own interest.

While affective computing systems in general often obscure how they infer information and predictions \cite{hauselmann2021fit}, the black-box nature of games is even more apparent. Games are often viewed as ``smoke and mirrors'' when it comes to the dichotomy between the game's parameter space and the conveyed aesthetic. Game designers are invested in hiding what lies within the game rules to facilitate a suspension of disbelief and thus make the experience more impactful and believable for players \cite{tence2010challenge}. Often game designers are relying on transgressive aesthetics \cite{jorgensen2019transgression} to create experiences that have a larger emotional weight. Although the area is still largely unexplored some games like Flying Mollusk's \emph{Nevermind} (2016) integrated computer models to guide their elicitation. In \emph{Nevermind} the game's transgressive aesthetics is amplified when the player is under stress and subdued when they calm down. Intentionally transgressive content---even when controlled by autonomous systems---can be tuned and managed more consciously; unintentionally harmful or offensive content, however, poses a much more complex problem. 

Of course some more obvious errors are easier to catch with simple pruning, but computational models can also encode biases---like gender biases or offensive stereotypes---which in turn result in harmful content \cite{sheng2019woman,lucy2021gender} unintended by the designer. A good encapsulation of the complicated nature of using AI generators, content moderation, and privacy is \emph{AI Dungeon} (Latitude, 2019), a text adventure game based on the GPT-2 \cite{budzianowski2019hello} and GPT-3 \cite{floridi2020gpt} language models. 
Latitude has become the focus of a controversy when they decided to take action against offensive content \emph{AI Dungeon}---mainly involving stories containing non-consensual sexual content and child pornography \cite{simonite2021aidungeon}.
Although much of the questionable content banned by Latitude was generated deliberately by their users, the model was also known to generate sexually explicit content seemingly unprompted, including ``writing children into sexual scenarios'' \cite{simonite2021aidungeon}. 
Users raised concerns about the decision of Latitude to address the issue with strict moderation, automatic flagging of problematic materials, and monitoring the content of users' privately generated stories. On one hand, this decision pushed all the \emph{responsibility} of the content to the users even though it was co-created with Latitude's algorithm; on the other hand, the human moderation of private content raised privacy concerns. 
The controversy showed that despite the best intentions of Latitude, a lack of \emph{transparency} and a clean line of \emph{responsibility} \cite{bostrom2014ethics} leads to detrimental outcomes for both the company and the end-users. The swift shift in how moderation was done on the platform made the already black-box system even harder to navigate for players, which in turn reduced both the transparency and the users' trust in the system. 
Many players felt unfairly flagged for content that either fell within community guidelines or was generated by the model virtually unprompted \cite{simonite2021aidungeon}. As Latitude was not the developer of the underlying foundation language model \cite{bommasani2021opportunities}, it was unclear how the system can be effectively \emph{audited} and since there was no established \emph{responsibility} for the system, all the blame fell to users who interacted with the model.

As we can see elicitation through AI-assisted systems has potentially harmful effects on the end user. On the one hand, video game companies can rely on affective computing models to fine-tune and personalise targeted monetisation strategies. This carries the danger of intentionally or unintentionally facilitating addiction or pressuring users on an emotional or social basis. On the other hand, generative systems can be unreliable and surprising and generate unwanted content without the designer's knowledge or the user's consent. To prevent subsequent issues, generative systems in games should demarcate a chain of responsibility for the model's output and offer tools to players to mitigate unwanted content. Even though foundation models offer a robust solution for generating the content, often the complexity, black-box nature, and ownership of the models limit the \emph{auditability} of these algorithms \cite{bommasani2021opportunities}. 

While leveraging predatory monetisation strategies has been a prevalent pattern---especially in the mobile games industry---not all studios have followed suit. A good counter-example is Six to Start, the developer of the immensely popular \emph{Zombies, Run!} (Six to Start) (2012) mobile exergame. The studio has not only forgone the usual dark design patterns, showing that games can be successful without putting psychological pressure on their players, but they have also been making a firm public stance against these practices \cite{hon2022you}.

\section{Sensing -- Privacy and Control}\label{sec:sensing}

Following elicitation, the next step of the affective loop is \emph{sensing}: the capture and processing of the manifested emotions. The central issue of sensing is that of privacy---i.e. when, how, and kind of data is being captured. While user privacy might define a clear issue in other domains, games present a special case. Particularly due to the interactivity of the medium, a certain amount of dynamic control is needed to maintain oversight over toxic and malicious actors in a game's ecosystem and prevent unintended or negative effects on the players' mental well-being. 

It has been shown that users of affective computing systems prefer clear notifications and potential control over the output of sensors \cite{reynolds2004affective}. This need is also supported by the \emph{principle of autonomy} when it comes to ethical applications \cite{cowie2015ethical}, and more concretely by the \emph{right of accurate portrayal} \cite{hauselmann2021fit}. Unfortunately, in multiple instances affective computing applications fail to address these needs as the data and data pipeline is kept secret from the user \cite{hauselmann2021fit}; game companies are no exception to this. In a recent comprehensive study on data types collected by the video game industry, Kr{\"o}ger et al. \cite{kroger2021surveilling} reveals that companies are collecting and inferring a wide range of data often without the user's knowledge.

Most of the data collection focuses on in-game behavioural metrics which can be used to predict player skill, preferences, or content consumption and spending habits. 
The inference of spending habits and personal identifiers---such as location, gender, financial status, etc---can clearly fuel harmful policies. Even game mechanics, which on the surface are there to benefit the player---such as matchmaking or dynamic difficulty adjustment---have led to concerns in the past. As mentioned in Section~\ref{sec:elicitation}, one of the most recent examples of dynamic difficulty adjustment working against the users was the case of EA Games. In a lawsuit, EA was accused to use difficulty adjustment to influence player spending on loot boxes \cite{valentine2020ealawsuit}. The lawsuit itself was later dismissed \cite{batchelor2021ealawsuit}, however, this case still highlights the public distrust towards systems that collect behavioural data.
This is not to say that games rely solely on such data. As Kr{\"o}ger et al. point out, game companies are adapting sensor data in their datasets at an increasing rate \cite{kroger2021surveilling}. Eye-tracking, voice data, GPS information, and peripheral signals from smart accessories can all be used to enrich game datasets and potentially reveal a wide variety of personal information about the user.
In today's interconnected world, it is becoming exponentially easy to triangulate and infer the identity of players---using their in-game data, username, GPS location, preferences, and distinct play patterns they use to interact with the game \cite{makarovych2018like,myers2021try}---to the point where it is questionable whether game data can be truly anonymised at all \cite{kroger2021surveilling}. 

Complex player profiles can be built on inferred gender, age, socioeconomic status, and interests, which can fuel harmful models exacerbating problem behaviour such as gambling and excessive spending. Beyond static high-level profiles it is also possible to infer the emotional state of users through keystroke patterns \cite{vizer2009automated}, voice \cite{kroger2019privacy}, or in-game behaviour \cite{shaker2010towards,el2016game}. Although the secrecy of the industry is a major concern, there is a tradeoff in terms of privacy and autonomy that is afforded to the players. Similarly to how behavioural analytics is used to infer a rich player profile \cite{bakkes2012player}, emotional data can also be used to model and subsequently enhance the play experience \cite{yannakakis2014emotion,hare2021player}. These types of tradeoffs were identified by Ishowo-Oloko et al. as the \emph{transparency–efficiency
tradeoff of human-machine cooperation} \cite{ishowo2019behavioural}. While not always applicable to human-computer interaction, there are instances where the inherent bias against AI \cite{rovatsos2019we} can hinder a human-computer system if total transparency is maintained.

\begin{figure}
    \centering
    \includegraphics[width=1.0\linewidth]{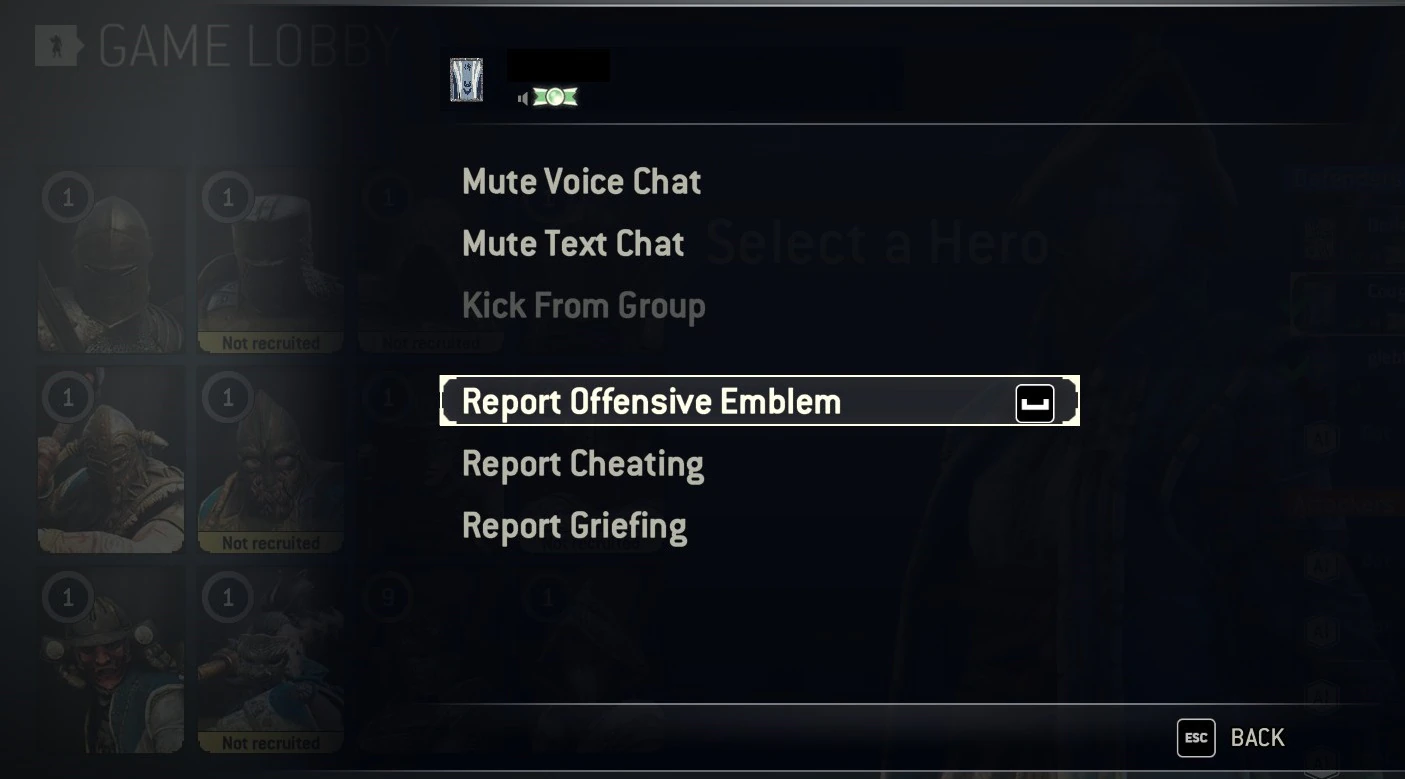}
    \caption{Traditional flagging methods for toxic behaviour in \emph{For Honor} (Ubisoft, 2017). Such methods often depend on user reporting which, in turn, may allow many toxic events to remain unnoticed and stay unreported.}
    \label{fig:toxic}
\end{figure}

Models incorporating emotional data can also be used to enhance other game systems involved in the moderation of user content and interaction. Most recently, Canossa et al. presented a robust method to flag the occurrence and severity of toxic and emotionally abusive behaviour in \emph{For Honor} (Ubisoft, 2017) \cite{canossa2021honor}. Although the input features of these models are mainly behavioural in nature, the inferred actions are emotional. While community guidelines are generally presented clearly, human moderation can become cumbersome with the breadth of data increasing with new players. Additionally, traditional reporting systems rely on user input, which could come with its own limitations including unreported events and subversion of the system by malicious users (see Fig.~\ref{fig:toxic}). As toxic players are often trying to find new ways to circumvent regulations, in the future automatic flagging systems---that incorporate emotional data as part of their input or output features---can make games a safer place for players. The aforementioned study is a good indicative example of how affective computing applications can be used to enhance the \emph{predictability} of the system towards its users and how we can use AI to both deliver clear value to players and improve the game experience.

When it comes to privacy and \emph{transparency} the state of practice in the game industry appears to be severe. The lack of consent, autonomy, or in many cases just knowledge about the collected data and its usage is a serious and prevalent ethical issue.
In many instances this ethical failing can be traced back to lax regulations, where legal requirements for consent and compliance are technically fulfilled but do not facilitate transparency in a tangible way.
However, despite recent legal efforts to institutionalise privacy requirements such as the GDPR of the EU, many industry players are still falling behind, unable or unwilling to address the most pressing issues \cite{kroger2019privacy}. 
On the other hand, not all data is collected for the purpose of exploiting and manipulating users. Often a wide range of data---otherwise considered ``non-essential''---can enable the automatic flagging and moderation of toxic and malicious users. Some players aim to actively harm others in- and outside of the game through emotional abuse. Affect-driven applications of AI can offer solutions for identifying both the occurrence and the impact of toxic behaviour, making online games a safer, more reliable and \emph{predictable} space for everyone involved. For this to happen, however, clear rules have to be put in place and users must be notified of what kind of data is being collected for what purpose. Even though the detection of toxic behaviour and negative gameplay outcomes might be beneficial for the players, it doesn't mean that the principle of \emph{transparency} cannot be upheld.

\section{Affect Detection -- Transparency in Limited Information Systems}\label{sec:detection}

The third core step of the affective loop is \emph{Affect Detection} which refers to the computational processing and prediction of certain aspects of affect \cite{calvo2010affect}. One of the major ethical challenges of deploying affective models, in general, is their \emph{transparency} towards users \cite{mikkelsen2017ethical,hauselmann2021fit}. More often than not companies are not disclosing that user data is modelled, let alone informing users about their system's predictions. Similarly to privacy challenges discussed in Section~\ref{sec:sensing}, the issue of transparency is of ambiguous nature too. 

We have already touched upon the issue of games as limited information systems in Section~\ref{sec:elicitation}. Games often withhold information to create uncertainty, decrease cognitive load, and construct challenges for players \cite{wang2011game,koster2013theory,abuhamdeh2015enjoying}. Moreover, games limit the up-front information that players have access to so that they facilitate learning \cite{koster2013theory} and an experience of \emph{flow} \cite{cowley2008toward}. Abuhamdeh et al. have shown that greater outcome uncertainty does indeed lead to greater satisfaction when the player is succeeding \cite{abuhamdeh2015enjoying}. They also found that as perceived competence rises, suspense and uncertainty become a major facilitator of \emph{intrinsic motivation} in video games; for an overview of intrinsic motivation in games see \cite{rigby2011glued}. Because uncertainty is a fundamental element of game design, it is very difficult to mitigate issues of \emph{transparency}. While explainable AI frameworks generally advocate for open communication towards the user about model predictions \cite{zhu2018explainable,zhu2021open}, when it comes to in-game adjustments this can be detrimental to the player experience \cite{ishowo2019behavioural}. It is important, however, that this tradeoff only applies to systems that use behaviour and emotion prediction to adjust in-game content, where the models only reach as far as the ``magic circle'' of the game experience \cite{huizinga1955homo,tekinbas2003rules}.

It is important to note that games do not exist in a vacuum and the experience is far from being a closed bubble. In systems where affect detection is used to inform monetisation strategies, users should be informed in a clear and comprehensive way about the output and the goal of the algorithm to preserve the system's \emph{transparency} and \emph{predictability}. Of course, as cited above, some game studios have a bad track record keeping users informed about their practices behind the curtains \cite{kroger2019privacy}. Although some of the secrecy can be chalked up to attempts to address the \emph{incorruptibility} of the system by keeping it obscure, more often than not it seems video game companies rather want to protect their resources (e.g. in terms of trade secrets, trained models, and datasets).

Beyond the questions of \emph{responsibility}, \emph{transparency}, and \emph{auditability} when it comes to detecting emotional and behavioural outcomes, game developers must also face the consequences of the inherent bias present in AI systems. For example, even though it might be responsible and beneficial to filter players based on certain emotional or behavioural states, models can also propagate unseen harmful biases. Even though \emph{transparency} becomes a critical duty of developers in such instances, other ethical standards must be drawn as well to preserve the integrity of these applications and reconcile with the experience being provided to the players. 

One of the most common causes attributed to algorithmic bias is a faulty dataset \cite{oniel2016weapons,lum2016predict,yapo2018ethical,gebru2021datasheets}. The more apparent issues in this regard are a skewed population, lack of control for diversity, and the non-critical capture of historical biases \cite{oniel2016weapons,lum2016predict}.
The latter of these issues makes it especially hard to mitigate algorithmic biases. On one hand, ``clean'', unbiased data might either not be available or impossible to attain. On the other hand, systems relying on historical biases can propagate patterns that seem true to a casual observer and are only revealed as biased under a more critical analysis \cite{roselli2019managing,gebru2021datasheets}. The issue becomes more severe due to the lack of \emph{transparency} and \emph{auditability} in the field. It is often next to impossible to recognise a biased dataset unless the algorithm breaches the trust of the users in a serious and very apparent way.
The most prolific of these instances are tied to sexist and racist outcomes \cite{yapo2018ethical,gebru2020race}. In one instance \emph{Google Photos'} algorithm was mislabelling pictures featuring black people as ``gorillas'' \cite{vincent2018google}. Of course, Google's algorithm was not created to be racist. The issue instead stems from the lack of diversity in the dataset that was used to train the model. As the model was trained on a dataset featuring predominantly white people, it learned to associate ``whiteness'' with ``people''. This error reveals a fundamental issue with a less-than-critical approach to data. Historic and institutionalised injustice defines our social reality. As injustice is ingrained at a fundamental level in our society, this type of bias is very hard to eliminate. The responsibility of the curators of large datasets and the developers of AI models is to apply critical forethought to processing and modelling to reduce the impact of these biases. 

While models used in the video game industry can arguably fail in similar contexts, there are more potential pitfalls unique to games. Similarly to other skewed datasets, game data can also be skewed towards atypical players, deriving an unfair distribution of the population as a whole. One example of this would be the overabundance of data from players with large amounts of playtime. Because they might not represent the population, an algorithm that focuses on these players would likely produce sub-optimal content for the remainder of the player base. This can especially be true in the initial phases of development as initial testers tend to be young and relatively good players. Another example of in-game bias would be discrimination towards atypical behaviour or emotional response. Systems monitoring toxic behaviour and bots are often based on high-level aggregated data on in-game actions and chat interactions \cite{canossa2021honor}. If these algorithms do not account for diversity and expect a behavioural and emotional response based on a Western, neuro-typical audience, they might flag good-faith players who are not conforming to certain behaviour or communication standards. On the other end of the spectrum, outliers can outperform the expectations of the model and are labelled as cheaters as happened to Julias Jackson, an autistic boy on the \emph{Xbox Live} ecosystem \cite{fahey2013xbox}. This error reveals the fragility of many automated systems when they have to apply their predictions outside of their trained boundaries. As we cannot be sure where those boundaries are in the wild \cite{mikkelsen2017ethical}, the challenge of organising and labelling our training data becomes a core issue. There is only so much we can anticipate in terms of future diversity requirements and even if we do, we might lack the tools to label our data correctly. Although it is relatively easy to rely on user reporting---especially when it comes to moderation---this also injects a large amount of bias into our systems both by malicious and good-faith users. Unfortunately, user reporting can often exacerbate other underlying biases, such as men accusing women of cheating for outperforming their peers \cite{ashcraft2016overwatch}. Even though these AI tools are in most cases used as flagging systems and not automated banning systems, the story of Julias Jackson shows that the infrastructure is far from being perfect, even with human oversight. 

Nevertheless, many industry leaders are well-invested in creating fulfilling experiences that promote player well-being. As part of these initiatives, companies such as EA and Nintendo \cite{johannes2021video}, Ubisoft \cite{melhart2019your} and Riot Games \cite{monge2022effects} often share data with academics and enable studies into fostering well-being, combating toxicity, and evaluating content moderation. Beyond their main purpose, these studies also expose some of the data that is collected and the information that is detected by these companies lending \emph{transparency} and \emph{auditability} to the industry. These types of cooperations provide a good example of how data transparency can bring value to industry players.

\section{Adaptive Systems -- Ownership in the Affective Loop}\label{sec:adaptation}

\emph{Adaptation} and affect expression are the steps that close the affective loop. Within games, an affect-based interaction system uses the parameterised output of its affect detection module to adjust the game's parameters to the user experience. The type of adaptation can occur at a macro level through an orchestrator that governs content in the game or the micro level through the behaviour and affect the expression of individual game agents. The goal of the game adaptation might differ depending on the given use case. It might serve to maintain, amplify or change the user's experience, but regardless of the interaction task, the adaptation module will yield a new set of emotional stimuli, thereby, closing the affective loop (see Fig.~\ref{fig:affective_loop}).

When we look at an adaptive system described by the affective game loop as a whole the question of ownership arises. While AI models can incorporate data from a large number of players, it is unclear how much ownership these players have over these affective models. This issue is even more complex in closed ecosystems that can facilitate co-creation with AI designers. As discussed earlier in Section~\ref{sec:related_work}, affective computing applications face challenges addressing these questions under the current regulatory frameworks \cite{jost2020two,hauselmann2021fit, kroger2023surveilling}. Even though users should have (at very least) rights to have control over their own data, to portray themselves accurately, to be forgotten \cite{villaronga2018humans}, and to be self-determined, in reality, information inferred from big data is often exempt from the same legal protection afforded to first-hand personal data \cite{wachter2019right}. To handle the data used to train models and the inferences made by these models on an individual basis Wachter and Mittelstadt propose the concepts of ``high-risk inferences'' and the ``right to reasonable inferences'' \cite{wachter2019right}. The former concept refers to inferences made from big data through algorithmic means that are either harmful to the privacy of the user or have low verifiability in what the authors call ``important decisions''---with loans and employment brought as examples \cite{wachter2019right}. The latter concept would enshrine a right that could force data controllers to provide certain information about their inferences:

\begin{quote}
    ``This disclosure would address (1) why certain data form a normatively acceptable basis from which to draw inferences; (2) why these inferences are relevant and normatively acceptable for the chosen processing purpose or type of automated decision; and (3) whether the data and methods used to draw the inferences are accurate and statistically reliable.'' \cite[page~8]{wachter2019right}.
\end{quote}

While this proposal---if it were to go into effect---could help provide more robust protection to users, unfortunately, when it comes to the models themselves matters get complicated. At the moment of writing, IP protection virtually takes precedence over the individual's autonomy over their personal data when it comes to the trained models themselves. As the models are considered only ``inferred from data'' they are further removed from the users whose data is used to create the models \cite{wachter2019right}. One key challenge with the existing legal framework is that the training data and the model are not as well separated or modular as the guidelines suggest. In many cases, it is possible to reverse engineer the model and extract information about the training set including sensitive personal information about the original subjects \cite{tramer2016stealing}. There are algorithmic solutions to address this problem, however. Ongoing research in the field of \emph{machine unlearning} aims to offer methods that attempt to remove knowledge from a trained model as if the given datapoint was never part of the training set \cite{guo2019certified,graves2020amnesiac,bourtoule2021machine,sekhari2021remember}. Although early approaches focused on very specific applications---such as \emph{decremental learning} in Support-Vector Machines \cite{cauwenberghs2000incremental}---recent methods seem to be able to generalise well over different architectures \cite{guo2019certified,bourtoule2021machine,sekhari2021remember}. 
However, for a practical application of \emph{machine unlearning}, the owner of the model has to retain the raw training data in most cases. Not all methods require this from a technical perspective \cite{sekhari2021remember}, but the removal of the datapoint has to be verified to preserve the \emph{predictability} of the system. Although the implementation of unlearning would most certainly pose a computational and organisational overhead, it is still more cost- and resource-efficient than retraining the models from scratch. Even though \emph{machine unlearning} has not been adopted widely yet, contemporary research results show a promising path ahead for mitigating some of the privacy issues in small-scale architectures.

Ethical questions in small-scale models can potentially be addressed through the aforementioned methods, however, far more complex ethical challenges are posed by large-scale world models or \emph{foundation models} \cite{bommasani2021opportunities}. These are large-scale pre-trained models built using hundreds of billions of parameters and massive-scale datasets often scraped from the internet indiscriminately. Foundation models have the potential to provide basic knowledge in a domain or generate content out of the box. While many applications like \emph{AI Dungeon} utilize these foundation models, the lines of \emph{responsibilities} are blurred as a given company has no access to the source code or the original data of these models. As the source of the data is often scraped from the internet, the ethics of constructing such a dataset is also highly questionable. As users are not notified they have no way to revoke their participation or retaliate against their creations used for training these models. In addition, the underlying data is often discarded or kept secret, removing even the moral right to the output of these generators once the model is constructed. Moreover, as these foundation models can only be constructed using an immense amount of resources, large industry players can essentially monopolise the market. This trend can already be seen in language-based models, where the current dominance of the closed-source GPT-3 and---increasingly more popular---GPT-4 \cite{openai2023gpt4} models imply that new applications have to subscribe to the black-box rules of that system. Ownership over the input, output and the models themselves is not a trivial problem.
The models are owned by their respective companies, and even though users generally retain rights to their input and to the output of the models, there are some major caveats. As an example, the Terms and Conditions of \emph{OpenAI} to their GTP-3 and GPT-4 algorithms\footnote{\url{https://openai.com/policies/terms-of-use}} which warns users that \emph{OpenAI} themselves retain the rights to both the user input and the system output to improve the system in the future. Nevertheless, users have input confidential information into the system that lead to security leaks. Most recently in the case of \emph{Samsung}, where confidential notes and source code has been leaked through ChatGPT \cite{maddison2023samsung}. While users can \emph{opt-out} of this data collection, it is unclear if already submitted data can be removed from the trained model.
The ownership issues are further complicated by the secretiveness of the industry stakeholders. 
Industry players are often invested in creating legal opacity around their systems through restrictive licences and digital rights management tools to restrict the transparency and usage of the software. Although this type of opacity does not stem from the AI architecture itself, it prevents public access to the inner workings of such systems and limits the overview of the AI decision-making process \cite{beckman2022artificial}.

\begin{figure}[!tb]
    \centering
    \includegraphics[width=1\linewidth]{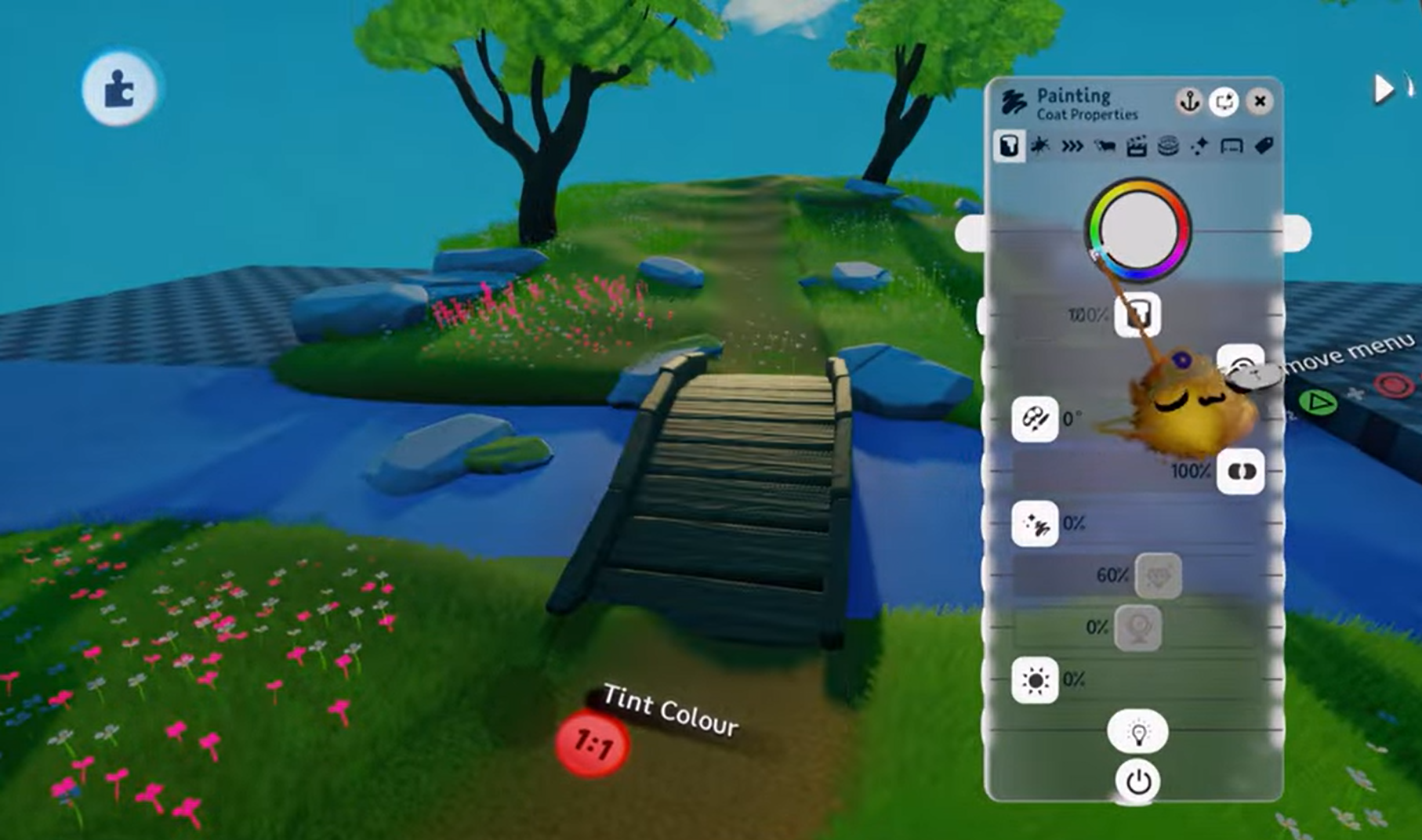}
    \caption{In \emph{Dreams} (Sony Interactive Entertainment, 2020), players are able to create their own experiences using a complex editor, reminiscent of the user interface of professional game engines.}
    \label{fig:editors:dreams}
\end{figure}

Even though the ownership over the models themselves is a central question, we must not forget about the ownership and responsibility over the output of said models either. Who owns the results of a human-AI co-creation process? Looking at contemporary legal frameworks, it is hard to say. Of course, judgement can be passed based on the circumstances and the particularities on a case-by-case basis, but there is no apparent clear line \cite{eshraghian2020human} and in most cases, the question is sidestepped entirely by the end-user licence agreement of specific AI-assisted tools. There is no comprehensive framework for either professional creative tools or interactive media meant for entertainment. The matter is further complicated because---even if we focus just on games---it can be hard to make a clear demarcation between a creative tool and curated entertainment. A recent example of this conundrum is Media Molecule's \emph{Dreams} (Sony Interactive Entertainment, 2020), a ``game about making games''---see Fig.~\ref{fig:editors:dreams}. Although \emph{Dreams} presents itself very much like a game, it perhaps has more in common with game engines, such as \emph{Unity}\footnote{\url{https://unity.com/}} or \emph{GameMaker}\footnote{\url{https://gamemaker.io/}}. Nevertheless, until recently games created with \emph{Dreams} were not monetisable by the creators and solely beholden to Sony's PlayStation ecosystem \cite{castello2020dreams}. While Media Molecule maintains that their users retain the rights to their own creations, the options for the users to exercise these rights remain limited\footnote{Media Molecule launched a Beta Evaluation for projects that aim to commercialise their creations off the PlayStation ecosystem. However, this program is not open to everybody and approval is granted on an opaque case-by-case basis. Interestingly---as of writing---games are completely excluded from the program. (Read more at \url{https://docs.indreams.me/en/community/news/dreams-beta-evaluation})}. After the runaway success of user-created content such as \emph{Defense of the Ancients}---originally created in the \emph{Warcraft III} (Blizzard Entertainment, 2002) map editor---which lead to a boom of Multiplayer Online Battle Area games it is easy to see why companies are trying to retain as much control over user-generated content as possible. There is a case to be made for the user's moral right over their creations, however. Especially in systems where the player demonstrates considerable creative effort during the co-creation process, they should be able to retain all rights to their own intellectual property. While contemporary examples are still subject to ad-hoc judgement, we expect the right to the output of co-creative systems to become a central topic in the near future as AI-powered generative systems become more ubiquitous. 

The conversation around co-creation is not just about rights but also responsibilities. Who is responsible for the output of the system when an agent learns to act like a bully, creates offensive and abusive content, or is instructed to generate misinformation? While it is easy---and companies are more than ready---to push the blame on the user, as we demonstrated before in Section~\ref{sec:elicitation}, addressing this issue is not as trivial. 
The main tools to combat the uncertainty around the models' output preemptively are \emph{transparency} and \emph{predictability}---not just from the AI perspective, but from the larger view of the organisation itself. A delineation between harmful content produced on purpose and as a result of a biased or erroneous algorithm has to rely on clear and transparent guidelines on how people are expected to interact with the system, and what the owners of the system consider malicious use. Predictability of the model output and the larger organisational response to adversarial attacks can facilitate a safer environment for all users involved.
The maintenance of the \emph{reliability} and \emph{incorruptibility} of the models should take precedence over the user's input. The \emph{responsibility} of designing and deploying these security measures should, however, fall on the creator of the system. Employing security measures in highly modular software systems is far from trivial given the integration of third-party models, especially out-of-box solutions. Nevertheless, as industrial systems maintain their opaqueness the end users cannot be considered fully autonomous. The extent of their autonomy will always be limited by the design of the application, the complexity of the system, and the limited transparency afforded to them.

The industry has been slow to react to the growing concern about trustworthiness; the landscape is changing, however. An excellent example of recent initiatives is Microsoft’s \emph{Xbox Transparency Reports}\footnote{\url{https://www.xbox.com/en-GB/legal/xbox-transparency-report}}. In this report, Microsoft publishes explanations and statistics about their content moderation policies to increase both the \emph{reliability} and \emph{transparency} of their ecosystem. Although the first report was just released in 2022, the company pledges to release these transparency reports every 6 months. If successful, a large industry player such as Microsoft can inspire the industry at large to follow suit.

\section{Other issues in game AI ethics} \label{sec:other}

There are several issues in the ethics of game AI that do not fit comfortably into our current structure, revolving around the affective loop. Even though these are not the core concern of our paper, and going in-depth on each question would require more space than we have here, we want to at least mention these concerns and provide some pointers for further reading.

All games are, in some way, partial representations of the world and processes therein. One might see games as being ``about'' certain real-world processes~\cite{bogost2010persuasive}. A key feature of games is also that you learn to perform the actions required to win the game as you play them; well-designed games are typically pedagogical sequences that introduce gradually harder versions of the same challenge, and it has been argued that this is a key component of why games are fun and appealing~\cite{koster2013theory}. While there are many games whose in-game processes are representations of peaceful real-life activities such as gardening in \emph{FarmVille} (Zynga, 2009) or warehouse stocking in \emph{Sokoban} (Thinking Rabbit, 1982), very many games represent some kind of real-life violence. Games about fighting in various forms are ubiquitous, and have been so since the birth of the medium; many, perhaps most, video games have ``hit'' or ``shoot'' as one of their most important mechanics. This might be because it is comparatively easy to design engaging games around fighting compared to any other purpose; regardless, combat is pervasive in games since the era of \emph{Pac-Man} (Namco, 1980) and its ghost-eating capability.

One potential issue of this is whether playing video games inspires, encourages, or teaches violent behaviour. There has been a debate around the effects of video games on violent behaviour for at least three decades, and many studies of varying quality. To some extent, this debate and field of inquiry can be subsumed under the broader question of media effects, where the effects of other media (such as TV) on violence and other behavioural aspects have been studied for a longer time. Although the idea that video games in some way lead to violent behaviour has some plausibility as video games teach some kind of skills, thorough and well-performed studies have largely failed to find any causal link~\cite{breuer2015violent}. Even if such an unknown link might be there, AI models can now assist in the isolation of such violence of toxic behaviours \cite{canossa2021honor}.

Given the very high computing demands of much current AI research and the advantages of having access to large datasets, it is worth pondering which modern game AI methods will benefit most. It is possible that modern game AI will exacerbate the divide between large developers with deep pockets, multiple titles and existing user bases and small independent game developers. If models trained on user data stay proprietary, the large developers will have a considerable additional advantage over small creators.

Concerns about fairness and bias are ubiquitous in machine learning, and as discussed previously, these concerns are very real for AI in games as well. It is often claimed that biased models come out of biased teams, in other words, that the composition of the human workforce defining and developing the AI solution impacts bias. This is certainly a concern in the game industry, which appears to be at least as demographically imbalanced as the rest of the tech industry \cite{bailey2021gender,passmore2017racial}.

Finally, a far-reaching potential ethical concern is that we one day develop artificial general intelligence, that is as capable (as we are) across a large range of areas and tasks. Games could have played a critical role in that development. Such entities might become very influential in human affairs, and may also gain the ability to improve themselves, potentially leading to what has been termed an \emph{intelligence explosion}~\cite{bostrom1998long}. If that happens, the alignment problem becomes acute: making sure that the goals and principles of such an entity are aligned with human society. Given that video games are abundantly used in AI research, it is worth pondering what impact training on video games might have on the ethics of a potential superintelligence. 

\section{Conclusions}\label{sec:conclusions}

This survey paper discussed thoroughly the various ethical aspects of artificial intelligence \emph{in} and \emph{for} games. We opted to view the most critical of these aspects under the affective game loop concept \cite{yannakakis2014emotion}. Based on that concept we reviewed the current game AI state-of-the-art and the game industry state-of-practice with respect to player experience \emph{elicitation}, \emph{sensing}, \emph{detection} and finally \emph{adaptation}. We raised a number of ethical dimensions and concerns and the current (lack of) measures and tools available to address them. We also made a number of recommendations and suggested future steps for making ethics an integral part of AI and games research and innovation. The dialogue between the game industry and academic stakeholders is currently active across the various conferences (e.g. GDC, IEEE CoG, FDG, CHI Play), seminars and summer schools (e.g. AI and Games Summer School) with a focus on the area. Moreover, the ethical aspects of AI in games and media at large are currently a top priority item in the agenda of policymakers (e.g. the European Commission) manifested through research and innovation projects\footnote{\url{https://www.ai4media.eu/}}\footnote{\url{https://learnml.eu/}} and policies such as the AI act\footnote{\url{https://artificialintelligenceact.eu/}}. We expect affective computing researchers to take a leading role in these efforts. Affective computing is uniquely positioned as a multidisciplinary field between sensor technology, AI, and applied psychology; hence it offers a comprehensive overview of most of the issues this survey has touched upon. Although in many cases the response to emerging issues has to be regulatory instead of technical, affective computing can still provide a shared language between the fields involved and help highlighting potential issues. This paper aims to further facilitate and moderate this dialogue among all stakeholders involved---AI and affective computing researchers and practitioners, game developers, and ultimately players---with the hope that ethical awareness is increased and that necessary action is taken for the mutual benefit of players and their games. 

\bibliography{citations.bib}
\bibliographystyle{IEEEtran}
\newpage
\begin{IEEEbiography}[{\includegraphics[width=0.9in,clip,trim={75px 0 75px 0},keepaspectratio]{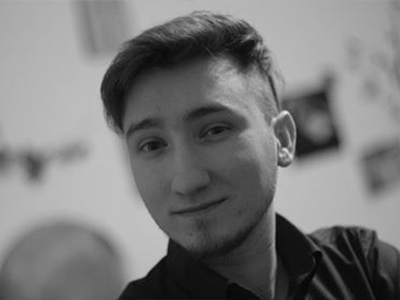}}]
{David Melhart} is a Senior Member of Technical Staff at modl.ai, and a Postdoctoral Researcher at the Institute of Digital Games, University of Malta. He received a MA degree in Cognition and Communication from the University of Copenhagen in 2016 and a Ph.D. degree in Game Research from the University of Malta in 2021. His research focuses on Machine Learning, Affective Computing, and Games User Modelling. He has been the Communication Chair of FDG 2020, a recurring organiser and Publicity Chair of the \emph{Summer School series on Artificial Intelligence and Games} (2018-2023), the Workshop and Panels Chair of FDG 2023, Editorial Assistant of the \emph{IEEE Transactions on Games}, Guest Associate Editor on the \emph{User States in Extended Reality Media Experiences for Entertainment Games} Special Issue of \emph{Frontiers in Virtual Reality and Human Behaviour}, and Review Editor of \emph{Frontiers in Human-Media Interaction}.
\end{IEEEbiography}
\vskip 0pt plus -1fil
\begin{IEEEbiography}[{\includegraphics[width=0.9in,clip,trim={40px 100px 40px 0},keepaspectratio]{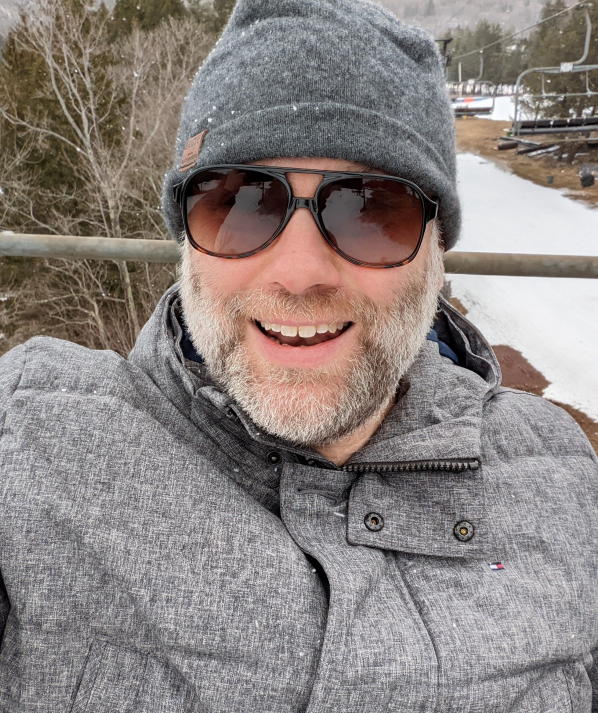}}]
{Julian Togelius} is a co-founder and Research Director of modl.ai, and an Associate Professor in the Department of Computer Science and Engineering, New York University. He works on artificial intelligence for games and on games for artificial intelligence. His current main research directions involve procedural content generation in games, general video game playing, player modelling, and fair and relevant benchmarking of AI through game-based competitions. Additionally, he works on topics in evolutionary computation, quality-diversity algorithms, and reinforcement learning. From 2018 to 2021, he was the Editor-in-Chief of the \emph{IEEE Transactions on Games}. Togelius holds a BA from Lund University, an MSc from the University of Sussex, and a PhD from the University of Essex. He has previously worked at IDSIA in Lugano and at the IT University of Copenhagen.
\end{IEEEbiography}
\vskip 0pt plus -1fil
\begin{IEEEbiography}[{\includegraphics[width=0.9in,clip,trim={75px 0 75px 0},keepaspectratio]{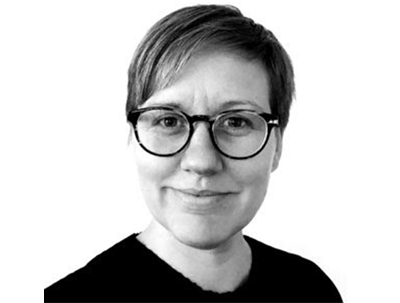}}]
{Benedikte Mikkelsen} is a co-founder and Chief Product Officer of modl.ai. She holds a BA degree in Architecture from The Royal Danish Academy of Fine Arts and an MSc in Media technology from the IT University of Copenhagen. Mikkelsen formed her first game development company Duck and Cover Games, focusing on learning game development, with Christoffer Holmgård in 2008, where they shipped multiple bespoke game titles for large public and private actors in Denmark. Subsequently, she worked with user experience and web application development, among other things developing design-driven integrated web and mobile platforms for actors in the public sector. She has experience working in research and development from international European Union Research projects focused on building non-expert end user behavioural modelling tools leveraging game telemetry data.
\end{IEEEbiography}
\vskip 0pt plus -1fil
\begin{IEEEbiography}[{\includegraphics[width=0.9in,clip,trim={75px 0 75px 0},keepaspectratio]{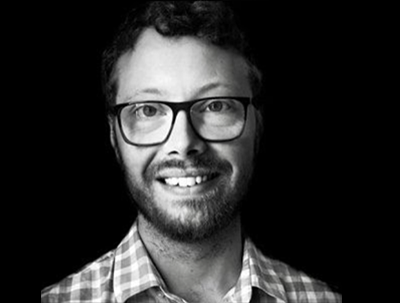}}]
{Christoffer Holmgård} is a co-founder and Chief Executive Officer of modl.ai. He holds a BA degree in Psychology from the University of Copenhagen and an MSc in Media technology from the IT University of Copenhagen. He worked in statistics, psychometrics, and organisational psychology with the Royal Danish Defence College from 2004 until 2011, maintaining the Danish national draft board intelligence test, conducting psychometric assessment of special forces operatives, fighter pilots, and officers, and supporting Danish veterans. In 2015, he earned a PhD in Artificial Intelligence and Procedural Content Generation from the IT University of Copenhagen and further earned a post-doctorate in Game Engineering from New York University. Before starting modl.ai, he served as a tenure-track Assistant Professor at Northeastern University and the head of their Master’s program in Game Science and Design. Holmgård formed his first game development company in 2008, Duck and Cover Games, focusing on learning games, together with Benedikte Mikkelsen. In 2011, he co-founded the award winning game studio Die Gute Fabrik, leading the studio as Managing Director, today serving as the chair of the board. During his tenure, the studio shipped multiple games on PC, Mac, mobile, and console platforms, securing more than 22 industry nominations and awards, including the GDC Innovation Award and the IndieCade Grand Jury Award.
\end{IEEEbiography}
\vskip 0pt plus -1fil
\begin{IEEEbiography}[{\includegraphics[width=0.9in,clip,trim={50px 0 100px 0},keepaspectratio]{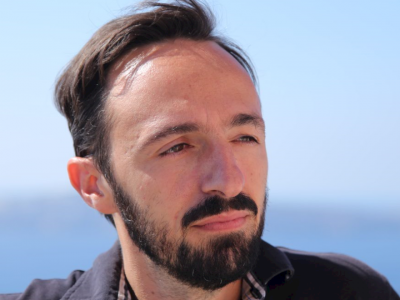}}]
{Georgios N. Yannakakis} is a co-founder of modl.ai, and a Professor and Director of the Institute of Digital Games, University of Malta (UM). He received the Ph.D. degree in Informatics from the University of Edinburgh in 2006. Prior to joining UM, in 2012 he was an Associate Professor at the Center for Computer Games Research at the IT University of Copenhagen. He does research at the crossroads of artificial intelligence, affective computing, games and computational creativity. He has published more than 350 papers in the aforementioned fields and his work has been cited broadly. His research has been supported by numerous national and European grants (including a Marie Skłodowska-Curie Fellowship). 
He is currently the Editor in Chief of the \emph{IEEE Transactions on Games}, an Associate Editor of the \emph{IEEE Transactions on Evolutionary Computation}, and used to be Associate Editor of the \emph{IEEE Transactions on Affective Computing} and the \emph{IEEE Transactions on Computational Intelligence and AI in Games} journals. 
Among the several rewards he has received for his papers he is the recipient of the \emph{IEEE Transactions on Affective Computing Most Influential Paper Award} and the \emph{IEEE Transactions on Games Outstanding Paper Award}. He is a senior member of the IEEE.
\end{IEEEbiography}

\end{document}